# Metastable Modular Metastructures for On-Demand Reconfiguration of Band Structures and Non-Reciprocal Wave Propagation

Z. Wu[1*], Y. Zheng[1, 2] and K.W. Wang[1]

[1] *Department of Mechanical Engineering, University of Michigan, Ann Arbor, MI 48109-2125*

[2] *State Key Laboratory for Strength and Vibration of Mechanical Structures, Xi'an Jiaotong University, Xi'an 710049, P.R. China*

*\* Corresponding author, email: wuzhen@umich.edu*

**Abstract**

We present a novel approach to achieve adaptable band structures and non-reciprocal wave propagation by exploring and exploiting the concept of metastable modular metastructures. Through studying the dynamics of wave propagation in a chain composed of finite metastable modules, we provide experimental and analysis results on non-reciprocal wave propagation and unveil the underlying mechanisms in accomplishing such unidirectional energy transmission. Utilizing the property adaptation feature afforded via transitioning amongst metastable states, we uncovered an unprecedented bandgap reconfiguration characteristic, which enables the adaptivity of wave propagation within the metastructure. Overall, this investigation elucidates the rich dynamics attainable by periodicity, nonlinearity, asymmetry, and metastability, and creates a new class of adaptive structural and material systems capable of realizing tunable bandgaps and non-reciprocal wave transmissions.

## 1   Introduction

Reciprocity of wave propagation is a fundamental principle [1] [2], describing the symmetry of wave transmission between two points in space. If wave can propagate from a source to a receiver, it is equally possible for the wave to travel in the opposite path, from the receiver to the source. Motivated by the concept of electrical diodes, directional flow of electrons in the presence of electric field, large amount of research attention have been devoted to explore the possibility of breaking the time-reversal symmetry and realizing one-way propagation in energy forms [3] [4] [5] [6] [7] [8] [9] [10]. Since linear structures alone cannot break the reciprocity in reflection-transmission if time reversal symmetry is preserved [11], considerable efforts have been devoted to realize non-reciprocal behavior in linear systems with additional symmetry breaking mechanisms. For instance, Fleury et al presented an acoustic circulator based on angular-momentum biasing through a circulating fluid [12]; Swinteck et al demonstrated bulk waves with unidirectional backscattering-immune topological states using superlattice with spatial and temporal modulation of the stiffness [13]; Wang et al proposed an all-optical optical diode via a "moving" photonic crystal to control the flow of light [14]; and Thota et al realized reconfigurable one-way acoustic wave propagation via origami folding with spatial modulated lattices [15]. In parallel to advances in spatiotemporal modulated linear systems, another major contribution in achieving non-reciprocal wave propagation is through nonlinear systems. Liang et al coupled a nonlinear medium with a supperlattice and accomplished unidirectional acoustic wave propagation by exploiting second-harmonic generation (SHG) of the nonlinear medium together with frequency selectivity of the linear lattices [7]. Boechler et al utilized the combination of frequency filtering and asymmetrically excited bifurcations in a defected granular chain to obtain rectification ratios greater than $10^4$ [8]. Popa and Cummer characterized an active acoustic metamaterial coupled to a nonlinear electronic circuit and demonstrated an isolation factor of >10 dB [16]. While many of these and other pioneer works pivoted primarily on the realization of unidirectional energy transmission, systems capable of on-demand tuning of non-reciprocal wave propagations, which are beneficial in many applications [17] [18], are yet to be addressed.

In complement but in contrast to previous contributions, in this research, we present a novel approach to accomplish non-reciprocal wave propagation with exceptional adaptivity by exploiting the concept of metastable modular metastructures, systems that exhibit coexisting stable states for the same topology. The proposed bottom-up metastructure concept is invested with direct pathways to facilitate global property adaptation by switching amongst the metastable states. Indeed, studies on metamaterials [19] [20], adaptive machines [21], and sensory adaptation systems [22] have all provided evidence that adaptivity may be induced by leveraging coexisting metastable states. To examine how the combination of nonlinearity, spatial asymmetry, periodicity and reconfigurability gives rise to unprecedented adaptive unidirectional wave propagation characteristics, the paper is organized as follows. In Sec II, the overall concept of the proposed unit module and metastructure is introduced. In Sec. III, the equations describing the nonlinear dynamics of the corresponding lattice chain are presented, followed by dispersion analysis of the linearized system. In Sec. IV, we present experimental evidence and analysis results of non-reciprocal wave propagation for the proposed architecture, discuss in detail the underlying mechanisms to realize non-reciprocal wave propagation and demonstrate adaptivity of such anomalous transmission afforded via tunable metastability.

## 2 Overall concept and example metastructure

The building block of the metastructure considered in this study is a metastable module consisting of a bistable spring and linear spring in series [23] [24]. A lab test stand is set up to explore the concept. The bistable constituent is generated by press fitting three magnets with repulsive polarization inside a 3D printed enclosure connected in parallel with a stabilizing spring realized via spring steel. Characteristic force-displacement profile of a bistable element is measured with an Instron machine, shown in Figure 1(a). The bistable element is then connected in series with a linear spring steel to form a metastable module. A characteristic force-displacement profile of the module is depicted in Figure 1(b). As shown in Figure 1(b), the building block exhibits a metastable range, where two metastable states (internal configurations) coexist with the same overall topology (global displacement). The experimental setup of the metastructure consists of a chain of such metastable modules connected in series horizontally, aligned with guiding rail and linear sliding bearings. Figure 1(c) depicts the top view of the experimental test bed and Figure 1(d) shows the corresponding schematic of the metastructure in which the bistable constituents are represented with buckled beams, the linear constituents are represented with coil springs and the inertial elements are symbolized with orange and yellow circles. As denoted in Figure 1(d), free length of the structure $L_{free}$ is defined to be the zero force position when all the bistable elements are buckled to the left and the global displacement z is defined as the additional deformation applied to the structure starting from the free length $L_{free}$ configuration. To investigate the non-reciprocal effect, two actuation scenarios are considered: one is forward actuation with actuator on the left side of the lattice chain and the other is backward actuation with actuator on the right hand side of the chain. The conceptual representation of the excitation scenarios are depicted in Figure 1(e). For illustration purposes, only inertial elements denoted by orange and yellow circles are presented in Figure 1(e) while the stiffness constituents connecting the masses are not shown. For both scenarios, displacement input $x_{in}$ is directly applied to the mass next to the boundary of the chain indicated by square, and output signal $x_{out}$ is measured one module away from the boundary marked with circle, Figure 1(c) and (e). During the experiments, both input and output displacements are measured with laser vibrometers.

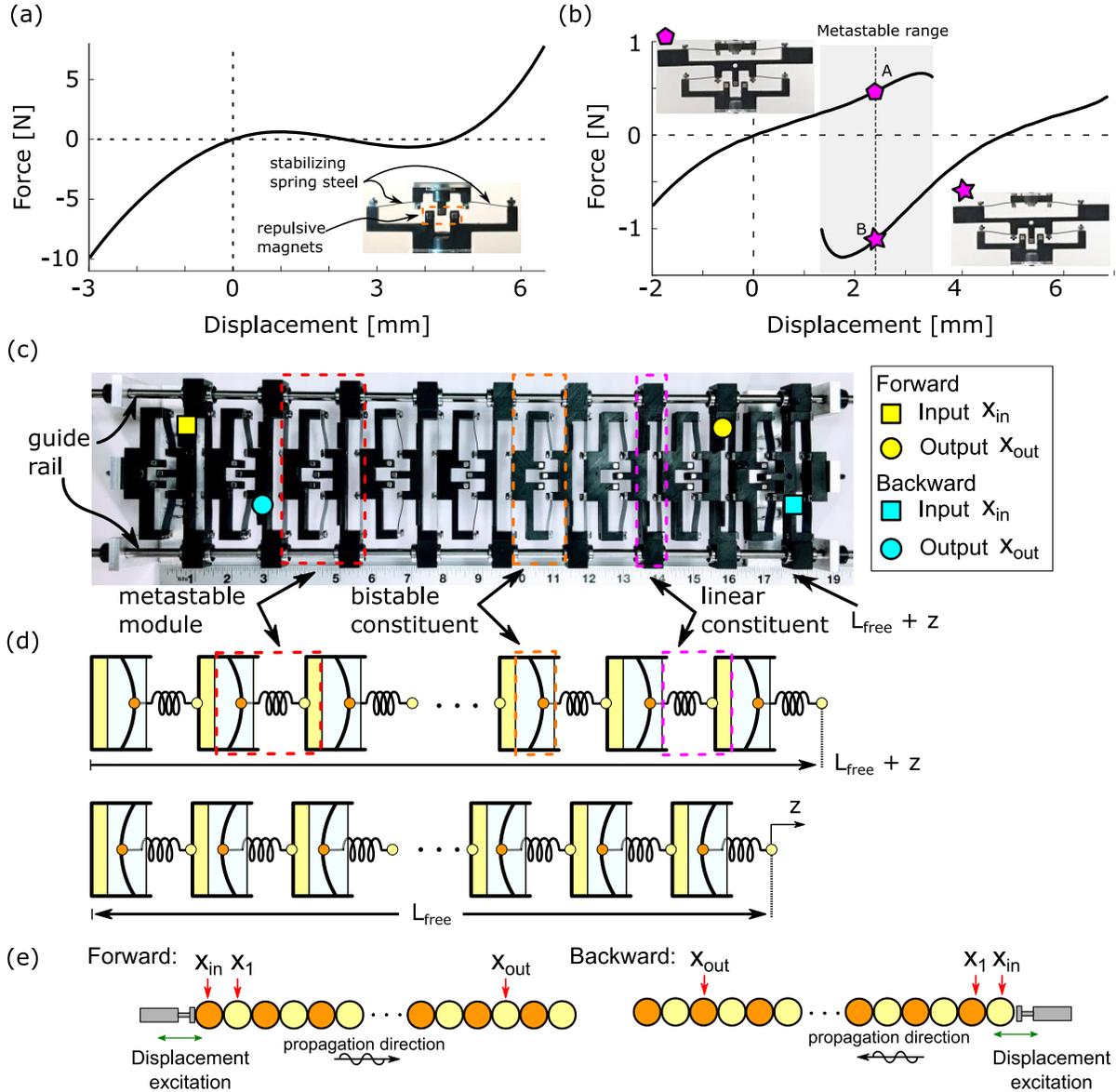

Figure 1. (a) Characteristic force displacement profile of a bistable element. (b) Characteristic force displacement profile of a metastable module. (c) and (d) Top view and corresponding schematic of the experiment setup. (e) Conceptual diagrams of metastable module assembled in series under forward (excitation from left) and backward (excitation from right) actuations.

## 3 Mathematical model and band structure analysis

### 3.1 Metastable states

Figure 2 depicts a 1D discrete lattice representation of N identical metastable modules connected in series as shown in Figure 1(c). Each metastable module, highlighted with red dashed box, consists of two masses $m_1$ and $m_2$ coupled via a linear constituent; the modules are inter-connected by bistable springs.

Without loss of generality, the bistable and linear restoring forces are assumed of the form, $F_{NL} = -k_1 x + k_3 x^3$ and $F_L = k_L y$, where $x$ and $y$ are the deformations of bistable and linear springs respectively. The total potential energy of the metastable chain for a fixed global displacement z measured from its free length $L_{free}$ can be expressed as:

$$U = \sum_{i=1}^{N-1}[\frac{k_L}{2}(x_{[i]2} - x_{[i]1})^2 - \frac{k_1}{2}(x_{[i+1]1} - x_{[i]2})^2 + \frac{k_3}{4}(x_{[i+1]1} - x_{[i]2})^4] + \frac{k_L}{2}(z - x_{[N]1})^2 - \frac{k_1}{2}x_{[1]1}^2 + \frac{k_3}{4}x_{[1]1}^4 \quad (1)$$

which is a function of internal mass displacements $x_{[i]1}, x_{[i]2}$, where subscript $i$ refers to the $i^{th}$ module. All internal displacements $x_{[i]1}$ and $x_{[i]2}$ are measured from the individual positions of the free length configuration. For a fixed global displacement z, the equilibrium positions of metastructure satisfies $\partial U/\partial x_{[i]1} = 0$ and $\partial U/\partial x_{[i]2} = 0$ under the constraint that $\sum_{i=1}^{N-1}(x_{[i]1} + x_{[i]2}) + x_{[N]1} = z$. According to the minimum potential energy principle [25], metastable states of the chain satisfy $\partial^2 U/\partial x_{[i]k}\partial x_{[j]l} > 0$, i.e. the Hessian matrix of the potential is positive definite.

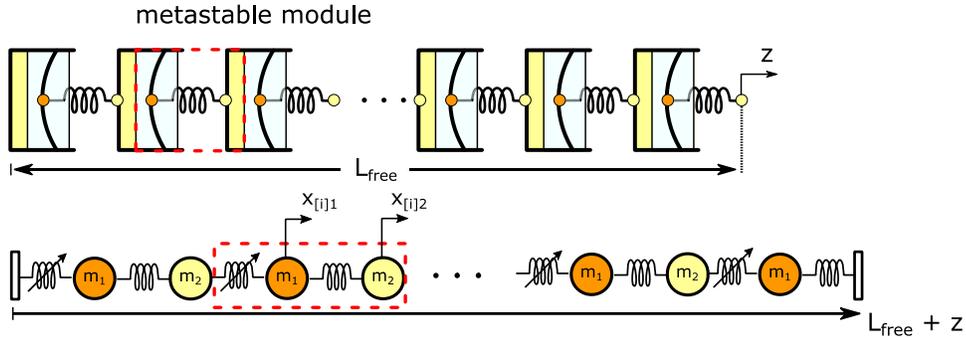

Figure 2. Schematic and discrete mass spring representation of an N metastable module assembled in series. A periodic unit cell, in this case same as the metastable module, is highlighted with red dashed box.

*3.2 Governing equations and linear dispersion analysis*

In general, for a fixed global displacement z, a chain of N metastable modules can have up to $2^N$ metastable states (internal configurations) [25] [26]. Starting from one of the metastable states, equations of motion for the $i^{th}$ module, can be expressed as:

$$m_1 \ddot{x}_{[i]1} + F_{NL}(x_{[i]1} - x_{[i-1]2}) + k_L(x_{[i]1} - x_{[i]2}) = 0 \quad (2a)$$
$$m_2 \ddot{x}_{[i]2} + F_{NL}(x_{[i]2} - x_{[i+1]1}) + k_L(x_{[i]2} - x_{[i]1}) = 0 \quad (2b)$$

Eq. (2a) is applicable to $\forall i = 2$ to N and Eq. (2b) is applicable to $\forall i = 1$ to N-1. Due to the fixed boundary conditions, equations of motion for first and last mass in the chain can be modified as

$$m_1 \ddot{x}_{[1]1} + F_{NL}(x_{[1]1}) + k_L(x_{[1]1} - x_{[1]2}) = 0 \quad (3a)$$
$$m_1 \ddot{x}_{[N]1} + F_{NL}(x_{[N]1} - x_{[N-1]2}) + k_L(x_{[N]1} - z) = 0 \quad (3b)$$

Depending on the excitation scenarios, Figure 1(c), external excitation will be directly applied to the first or last mass in the chain.

To establish a linear dispersion relation, we first linearize the equations of motion about its metastable state and the band structure is determined by modeling a repeating periodic unit cell of an unforced, infinite chain. For the diatomic chain depicted in Figure 2, the periodic unit cell is the same as a metastable module, and the linearized equation can be written as:

$$m_1 \ddot{\zeta}_i + \tilde{k}_{NL}(\zeta_i - \eta_{i-1}) + k_L(\zeta_i - \eta_i) = 0 \tag{4a}$$

$$m_2 \ddot{\eta}_i + \tilde{k}_{NL}(\eta_i - \zeta_{i+1}) + k_L(\eta_i - \zeta_i) = 0 \tag{4b}$$

where $\zeta_i$ and $\eta_i$ are small perturbation of mass $m_1$ and $m_2$ around its initial positions and $\tilde{k}_{NL}$ is the corresponding linearized stiffness of the bistable spring. Assuming solutions in the form of a traveling wave, i.e, $\zeta_i = A exp[j(\omega t - kiL)]$ and $\eta_i = B exp[j(\omega t - k(i+1)L)]$, where $k$ is the wave number and $L$ is unit length, the model is reduced to a standard eigenvalue problem:

$$\begin{bmatrix} (k_L + \tilde{k}_{NL})/m_1 & -(\tilde{k}_{NL} + k_L e^{-jkL})/m_1 \\ -(\tilde{k}_{NL} + k_L e^{jkL})/m_2 & (k_L + \tilde{k}_{NL})/m_2 \end{bmatrix} \begin{bmatrix} A \\ B \end{bmatrix} = \omega^2 \begin{bmatrix} A \\ B \end{bmatrix} \tag{5}$$

The band structure can then be determined by sweeping the wave number $k$ from 0/L to $\pi/L$. In general, due to the existence of multiple metastable states for a chain of N metastable modules, depending on the initial configuration, periodic repeating cell should be identified and similar dispersion analysis can be carried out accordingly.

*3.3 Analysis results of band structures*

For exploration purposes and without loss of generality, parameters used in the analysis are chosen to be of arbitrary unit. With $k_1 = 1, k_3 = 1, k_L = 0.2$，Figure 3(a) depicts the resultant force profile of the metastructure as global displacement changes. Due to a synergistic product of assembling together metastable modules, different internal configurations can be afforded via transitioning amongst these metastable states for the same global displacement z [23] [24]. Configurations A and B in Figure 3(a) for instance correspond to two periodic metastable chains of the same global length but with different internal configurations or metastable states. Periodic unit cells for the two configurations are highlighted with red dashed box in Figure 3(a) and the corresponding band structures for configurations A and B are depicted in Figures 3(b) and (c) respectively. For configuration A, the periodic cell is the same as a metastable module while for configuration B, the periodic cell consists of two metastable modules, Figure 3(a). As illustrated in Figure 3, property programmability can be realized via metastable state switching, demonstrated by the adaptation of band structures of the metastable chain. For configuration A, the two passbands are [0, 0.633] and [2, 2.098] whereas for configuration B, the first passband remains to be the same [0, 0.633] while the second passband shifts to [1.069, 1.242].

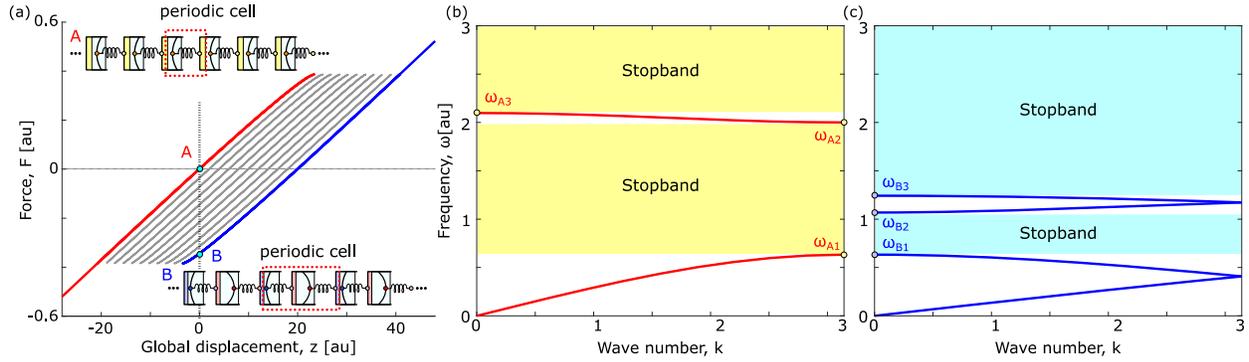

Figure 3. (a) Reaction force profile as a function global displacement for a 10 module metastable chain. Points A, B are two different configurations for the same global topology realized by internal configuration switching. (b) and (c) Corresponding band structures for configurations A and B. Passband are within [0, $\omega_{A1}$], [$\omega_{A2}$, $\omega_{A3}$] and [0, $\omega_{B1}$], [$\omega_{B2}$, $\omega_{B3}$] respectively, demonstrating the massive band structures tunability via internal configuration switching.

## 4    Non-reciprocal wave propagation and adaptation

### *4.1 Experimental investigation*

With global confinement of 45.1 cm (17.75 inches), two configurations I and II of the experiment setup are depicted in Figure 4(a). The two configurations are obtained by switching between metastable states and the periodic repeating cell is highlighted with dashed boxes. To first identify the band structures, a frequency sweep tests are performed for both configurations with input RMS acceleration 0.3 m/s$^2$ via an electromagnetic shaker. Due to limitation of the shaker, the frequency range is selected to be from 5Hz to 50Hz and the sweep rate is 0.05 Hz/s. Figure 4(b) depicts the displacement frequency response function (FRF) for forward actuation, with blue and red representing configurations I and II, respectively. As depicted in Figure 4(b), the first passband for configuration I extends from 5Hz to approximately 10Hz whereas for configuration II, the first passband lasts up to 18.5Hz, exemplifying the bandgap adaption as the metastructure is reconfigured. Additionally, such adaptation can be exploited in-situ. For instance, as depicted by the displacement time history in Figure 4(c), with input RMS displacement 0.2mm and input frequency 15Hz (gray). Starting with configuration II, the vibration energy is able to propagate through the chain since the excitation frequency is inside the passband (blue). We then switch to configuration I by sequentially reconfiguring two bistable constituents, depicted by the consecutive spikes in time history (black) in Figure 4(c). In this case, the band structure shifts from the passband to stopband, and therefore the output displacement is reduced significantly to nearly zero (red).

To illustrate the non-reciprocal wave propagation, the metastructure in configuration I is excited at 15Hz with increasing input amplitude for both forward and backward actuation scenarios. The excitation frequency is chosen such that it is inside the stopband of the linearized structure. Figure 4(d) depicts experimental results of transmittance ratio as input amplitude increases. The transmittance ratio (TR) is defined as the ratio of output RMS displacement over input RMS displacement $TR = |\bar{x}_{out}|/|\bar{x}_{in}|$, with solid and dashed lines representing forward and backward actuation scenarios, respectively. As indicated in Figure 4(d), with small excitation level, due to the bandgap effect, the output displacement is negligible compared to the input amplitude. As input amplitude increases to 0.3mm, the transmittance ratio of the backward actuation increases significantly while it remains to be low for forward actuation, as illustrated in Figure 4(d), providing experimental evidence on the start of non-reciprocal wave propagation. This unidirectional energy transmission phenomenon ends at input amplitude 0.58mm, at which sudden increase of the output amplitude for the forward actuation is also made possible. Such amplitude dependent wave transmission features corroborate with the studies on supratransmission, in which energy of a signal with input frequency in the stopband is transmitted through the chain as input amplitude exceeds certain threshold [27] [28] [29] [30]. Therefore, through intelligently integrating the supratransmission property of a nonlinear periodic chain with spatial asymmetry, the metastable structure is capable of attaining supratransmission at different excitation level depending on the excitation and propagation directions, enabling non-reciprocal wave propagation.

In addition to non-reciprocal wave propagation, endowed with metastability, the proposed structure is also capable of exhibiting onset of supratransmission at different excitation level by internal reconfiguration. As illustrated in Figure 4(e), the metastructure is excited under forward actuation at 20Hz for both configurations I and II. Since the excitation frequency is within a stopband for both configurations, under small excitation level, energy does not transmit through the chain in both cases. As input amplitude increases to 0.06mm, energy starts to transmit through the chain for configuration II whereas energy transmission is still prohibited for configuration I until excitation level reaches 0.45mm, after which wave propagates through the chain for both configurations. Such adaptivity on the onset of supratransmission for different configurations is crucial to create systems with on-demand tuning of non-reciprocal wave propagation characteristics.

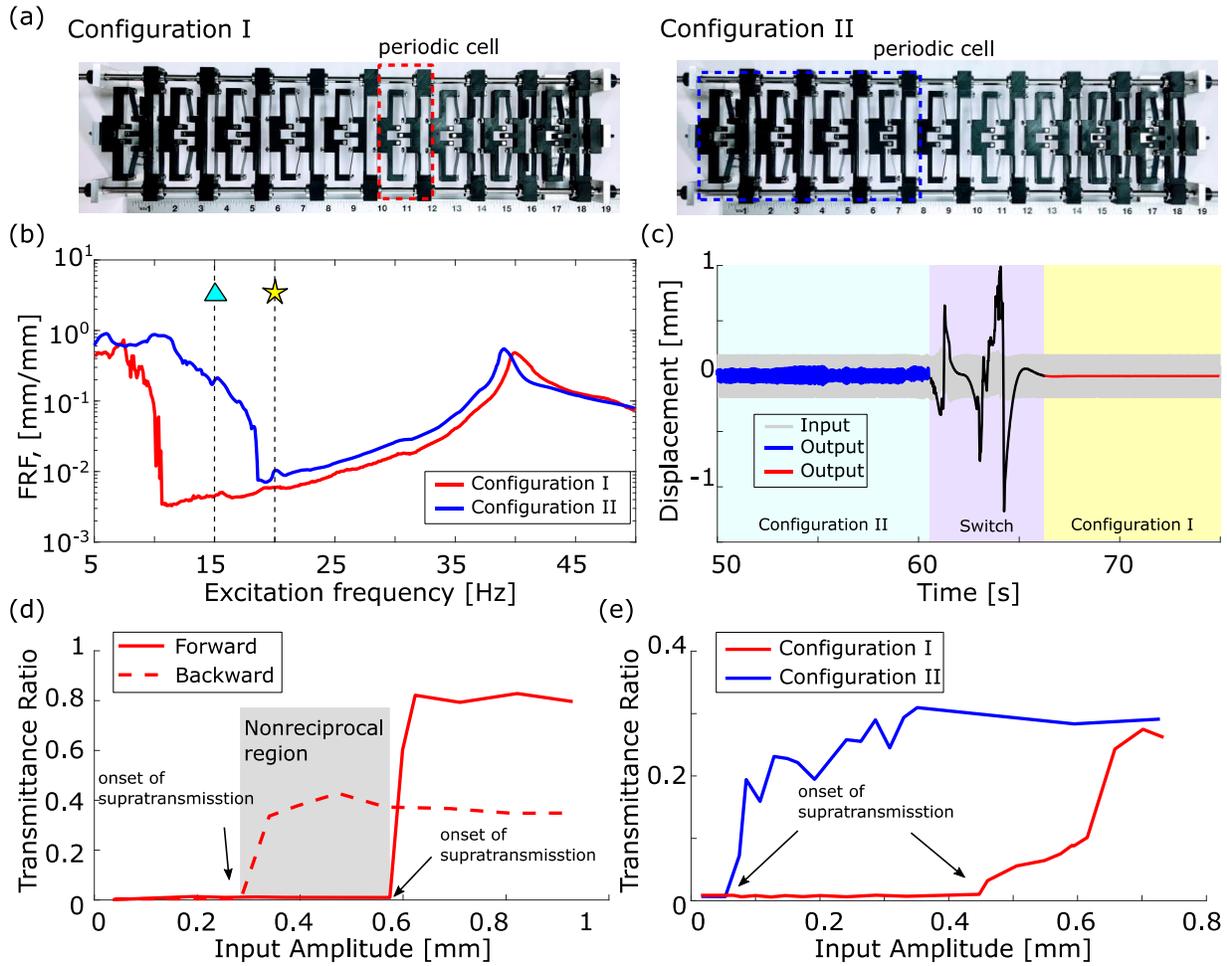

Figure 4.(a) Top view of experiment setup with two different configurations. Two configurations can be reconfigured by switching between metastable states. Periodic cell of each configuration is highlighted with dashed box. (b) FRF of output displacement over input displacement as input excitation frequency changes for configuration I (red) and configuration II (blue) with small excitation level. Frequency 15Hz (triangle) is in the passband for configuration II and stopband of configuration I and 20Hz (star) is in the stopband for both configurations. (c) Time history of output displacement with input RMS displacement 0.2 mm and input frequency at 15Hz. Signal changes in-situ from propagating to nonpropagating as configuration changes from configuration II to I. (d) Transmittance ratio (TR) for forward (solid line) and backward (dashed line) actuation for configuration I with input frequency at 15Hz. Shaded area denotes the region of non-reciprocal wave propagation. (e) Transmittance ratio (TR) for configurations I and II under forward actuation at 20Hz, demonstrating the adaptiveness of supratransmission.

*4.2 Analysis results – generation of non-reciprocal wave propagation*

To understand the mechanisms of non-reciprocal wave propagation of the proposed system, a detailed numerical analysis is performed using the system parameters in the dispersion studies discussed in Sec. 3.3 with masses $m_1 = m_2 = 1$. Wave propagation characteristics are first explored for configuration A, defined in Figure 3, under both forward and backward actuations. Small damping coefficients $\zeta=0.001$ is applied between lattices and simulations are performed for sufficiently long time (30000s) to reach steady state. Figure 5 depicts the displacement time history and FFT of corresponding velocities for input $x_{in}$ (magenta), output $x_{out}$ (cyan) and response of internal mass adjacent to input $x_1$ (gray), Figure 1(c). Driving frequency $\omega = 1.15$ is chosen to be within the attenuation band $[0.633, 2]$ of the metastructure for configuration A from dispersion analysis and two input amplitudes $\delta = 0.1$ and $0.5$ are considered.

For small input amplitude $\delta = 0.1$, despite the large discrepancy on steady state amplitude of internal mass (gray) between backward and forward actuations, the output response amplitudes (cyan) are both negligible compared to that of the input (magenta), Figure 5(a) and (c). Difference in response amplitudes of the internal masses is due to the inherent spatial asymmetry of the metastable chain introduced by modules with different elastic constituents. With finite lattice length, one end of the chain is grounded via a linear constituent whereas the other end is fixed through a nonlinear constituent, Figure 1. To understand the wave propagation characteristics, FFT of the corresponding velocities are depicted in Figure 5(b) and (d). Red dashed lines are band structure boundaries $\omega_{A1}$, $\omega_{A2}$ and $\omega_{A3}$ determined from the linear analysis, Figure 3(b). For both actuation scenarios, majority of the energy is localized around the fundamental driving frequency $\omega_d$ as well as its higher harmonics $2\omega_d$ and $3\omega_d$, Figures 5(b) and (d). In complement to the time domain analysis, the FFT results also reveal that response amplitude of the internal mass for backward actuation is orders of magnitude greater than that for the case of forward actuation, due to the inherent spatial asymmetry. In fact, amplitude of the second harmonic $2\omega_d$ under backward actuation is still two times larger than the amplitude of fundamental harmonics $\omega_d$ under forward actuation. However, since these dominant frequencies reside within the non-propagating zone of the metastable lattices, majority of wave energy does not propagate through the metastable chain. As exemplified by the frequency response of output signal for both scenarios, dominant frequencies of the remaining wave energy are inside the passband and prominently within the first passband. Yet, since the vibration amplitudes of the output signals are considerably reduced compared to that of the input, transmittance ratios are negligible.

As input amplitude increases to $x_{in} = 0.5$, forward propagation of wave is still largely prohibited, similar to the low amplitude actuation case, Figures 5(e) and (f). However, comparing Figures 5(e) and (g), the response of output signal increases substantially for backward actuation. More specifically, with large enough input amplitude, the subsequent internal masses (gray) instantly undergoes large amplitude vibration, in this case chaotic response, Figure 5(g). Therefore, even though the driving frequency is within the non-propagating zone, input frequency is immediately redistributed along a broad frequency band and wave energy is transmitted with frequencies spectrum inside the propagating passband, Figure 5(h). This frequency conversion property is a classical nonlinear phenomenon and also substantiates previous researches where supratransmission is enabled via nonlinear instability [27] [28] [29] [30]. Hence, we can conclude that non-reciprocal wave transmission with the proposed metastructure is facilitated through the interplay of spatial asymmetry, nonlinearity and periodicity. Use the same definition of transmittance ratio (TR), it is determined that with the given parameters, TR for forward and backward actuations are 0.01 and 3.73 respectively, demonstrating more than 2 orders of magnitude increase in transmittance as excitation direction changes and providing clear evidence of non-reciprocal wave transmission.

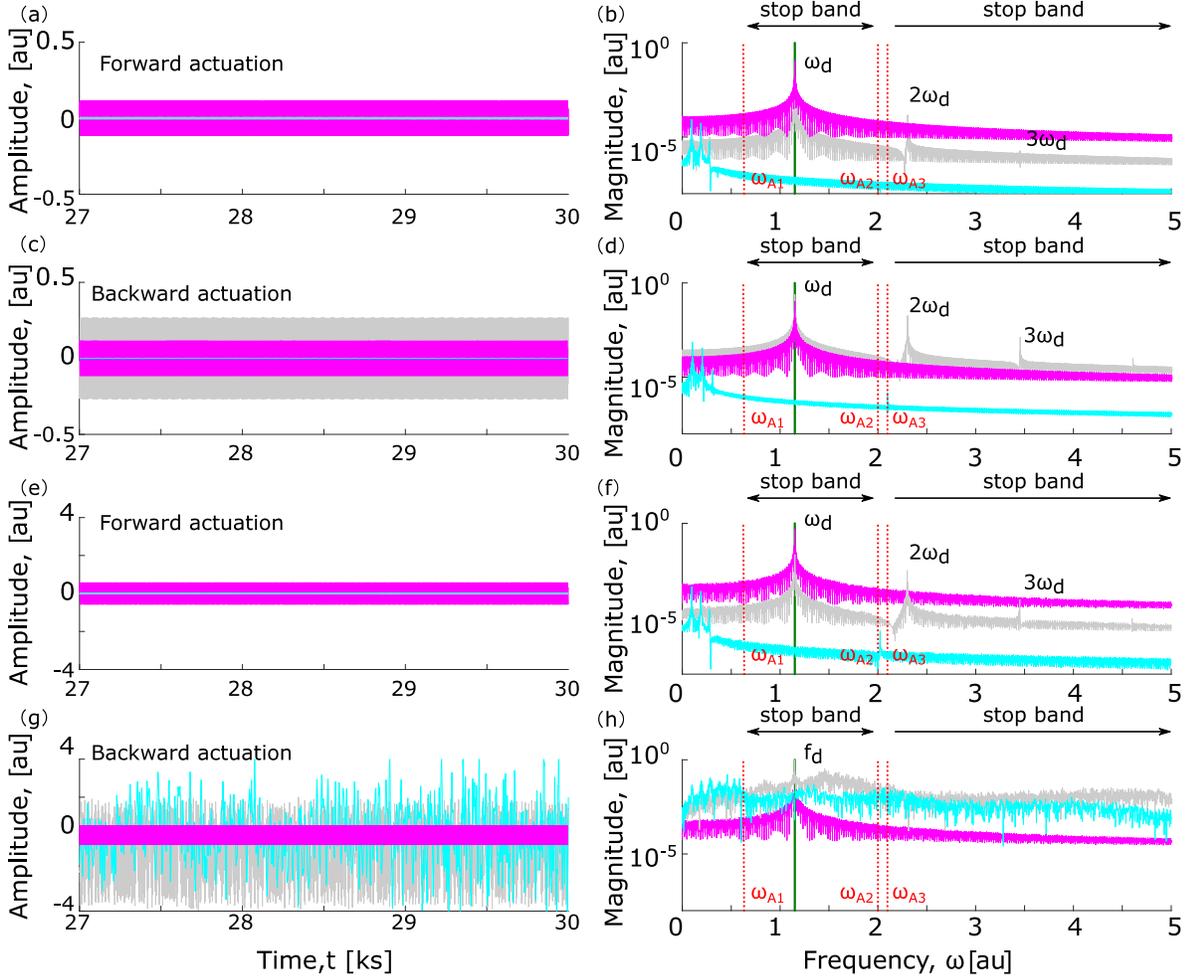

Figure 5. Steady state displacements of input (magenta), output (cyan) and internal (gray) masses for configuration A shown in Figure 3 under forward ((a) and (e)) and backward ((c) and (g)) actuation with excitation frequency $\omega_d = 1.15$ and excitation amplitude $\delta = 0.1$ ((a) and (c)) and $\delta = 0.5$ ((e) and (g)). (b), (d), (f) and (h), frequency domain analysis of corresponding velocities. Red dashed lines are band structure boundaries frequencies for configuration A predicted from linear analysis and green solid line is the input driving frequency.

## 4.3 Analysis results - adaptive wave propagation

Experimental and numerical studies discussed in the previous sections provide convincing evidence on the non-reciprocal wave transmission characteristics as input excitation amplitude varies. Experimental observations also indicate that such anomalous energy flow phenomenon is not only a result of the nonlinearity and spatial asymmetry of metastructure, but also tightly related to the bandgaps of the periodic chain. To further investigate the influence of internal reconfiguration (change of metastable states) on the transmittance ratio (TR) and non-reciprocity of the metastable lattice, numerical analysis is performed on the same metastructure by varying both internal configuration and actuation scenarios.

Figure 6(a) depicts the transmittance ratio as input amplitude changes under constant input frequency $\omega = 0.95$. This frequency is within the stopband for both configurations A and B, Figure 3.

For small input amplitude, configuration A (red lines), both forward and backward actuations have small transmittance ratio, similar to previous observations. Yet, due to spatial asymmetry, as input amplitude increases to 0.3 m, transmittance ratio for backward actuation increases significantly by 2 orders of magnitude, red dashed line, whereas TR for forward actuation remains to be low, indicating the start of non-reciprocal wave transmission. Same as the experimental observation in Figure 4(d), backward actuation is able to reach the onset of supratransmission with smaller excitation amplitude than forward actuation.

As input amplitude increases to 1.7, input energy is sufficient to trigger large amplitude vibration for forward actuation and is reflected by the large increase in transmission ratio, red solid line. Further increasing input amplitude beyond 1.7, wave energy will transmit in both directions. Hence, for configuration A, non-reciprocal wave transmission occurs for input amplitudes between 0.3 to 1.7. As the metastructure is changed to configuration B (bottom row in Figure 6(a)) by switching the metastable states (Figure 3(a)), with the same excitation frequency $\omega = 0.95$, amplitude range for such unidirectional energy transmission now shifts to between 0.1 to 0.2. This is due to the fact that comparing with configuration A, configuration B corresponds to a softer state, i.e. equilibrium position is at a shallower potential well, evidenced by a lower passband, Figure 3(b). Therefore, nonlinear instability is more readily attainable compared to configuration A. This demonstrates the adaptivity of non-reciprocal wave transmission characteristics as switching amongst the metastable states.

Additionally, as shown in Figure 3, alternating internal configurations can greatly affect the bandgaps of the metastable lattice, which is shown to be pivotal in manipulating frequency spectrum of the output signal, Figure 5. Hence, the effect of input frequency on the non-reciprocal wave transmission characteristics as switching amongst metastable states is investigated. Figure 6(b) illustrates the transmittance ratio as input frequency changes with constant input amplitude $\delta = 0.5$. It can be seen that for configuration A, non-reciprocal transmission exists for frequencies between [0.7, 1.7] and [2.3, 3] and changed to [1.7, 2] and [2.5, 3] as switched to configuration B.

To further explore the adaptivity of wave propagation characteristics over a wide spectrum of input parameters, Figure 6(c) depicts the transmittance ratio for forward and backward actuations with both configurations A and B as input frequency and amplitude varies. The transmittance ratio heat map shown in Figure 6(c) are in log scale with lighter color region corresponds to larger TR values. As demonstrated in Figure 6(c), for both configurations A and B, wave propagation characteristics for forward and backward actuations are considerably different for various combinations of input parameters. In fact, with the given system parameters, most of the signal transmit through the chain for backward actuation, whereas for forward actuation, wave energy does not propagate for some combinations of input frequency and amplitude in the parameter space, indicated by a greater area of dark space in Figure 6(c). More intriguingly, as configurations switches from A to B, significant adaptation of wave propagation characteristics can be observed for forward actuations, Figure 6(c). For instance, with input level $\delta = 1$ and frequency $\omega = 1.5$, initially blocked wave energy for configuration A can propagate through the chain as switched to configuration B. It's also worth noting that the non-propagating zone for configuration B, indicated by the dark blue area is much smaller compared to that for configuration A, corroborating with the previously discussed fact that configuration B corresponds to a softer state. From these observations, we can conclude that the proposed metastable metastructure is invested with massive adaptivity of non-reciprocal wave propagation characteristics by switching amongst the metastable states.

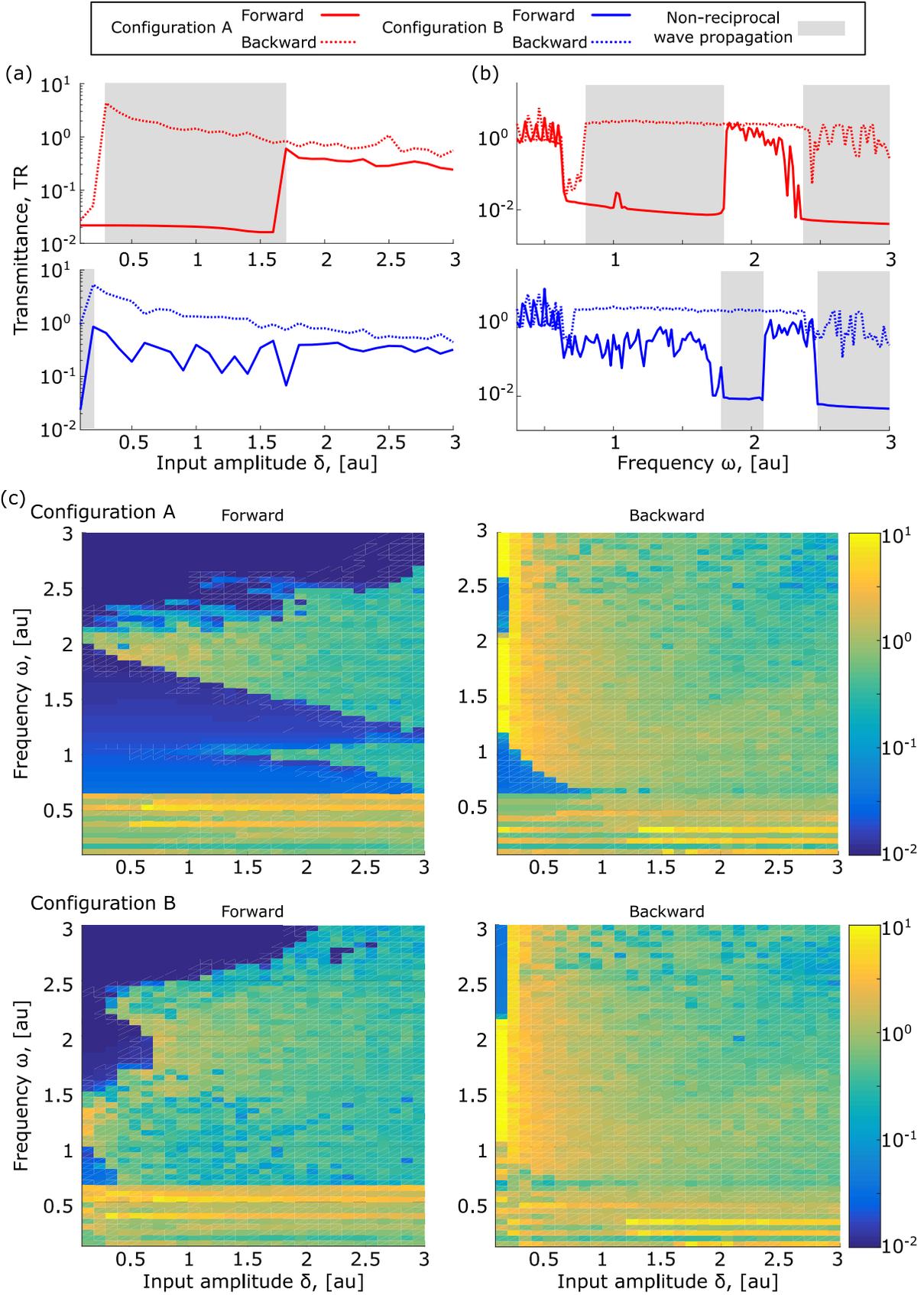

Figure 6. (a) and (b) Transmittance ratio (TR) for metastructure under forward (solid line) and backward (dashed line) actuation for configuration A (red) and B (blue) with shaded areas denoting parameter space for non-reciprocal wave propagation. (a) Varying input amplitude with fixed frequency $\omega_d = 0.95$ and (b) Varying input frequency with fixed amplitude $\delta = 0.5$. Gray area indicates the region of non-reciprocal wave transmission. (c) Contour plot on transmittance ratio (vs. input frequency and amplitude) of forward and backward actuation for configurations A and B.

## 5 Conclusion

In this paper, we present a novel approach to achieve the on-demand adaptation of band structures and non-reciprocal wave propagation. We found that unidirectional energy transmission can be facilitated by triggering the onset of nonlinear supratransmission at different excitation levels for different actuation directions due to spatial asymmetry. Therefore, by intelligently leveraging nonlinearity, symmetry, periodicity, and reconfigurability, one can tune and control wave propagation via a metastable modular metastructure. By exploiting the programmable properties invested in such modules and structures, we have developed unprecedented adaptivity of wave propagation characteristics with internal reconfiguration and metastable state changes, as input amplitude and frequency varies. Results of the present work will pave the way for further analytical, numerical, and experimental studies of adaptive non-reciprocal wave propagation in higher-dimensional systems. Lastly, since the approach depends primarily on scale-independent principles, it could foster a new generation of reconfigurable structural and material systems with unconventional wave characteristics that are applicable to vastly different length scales for a wide spectrum of applications.

**Acknowledgment**

The authors gratefully acknowledge the support of the U.S. Army Research Office under grant number W911NF-15-1-0114. Yisheng Zheng would also like to acknowledge the support from China Scholarship Council.